\documentclass[aps,prl,singlecolumn,nopacs,superscriptaddress,longbibliography,reprint]{revtex4-1}

\usepackage[breaklinks=true,colorlinks,citecolor=blue,linkcolor=blue,urlcolor=blue]{hyperref}
\usepackage{epsfig,mathrsfs,color,latexsym,subfigure,marginnote,graphicx,verbatim,relsize,mathrsfs,color,array,amsmath,amsfonts,amssymb,graphicx}
 \usepackage{multirow}
 \usepackage{makecell}
\usepackage{lipsum}

\renewcommand{\BibitemShut}[1]{}

\newcommand{\rpl}[1]{\textcolor{black}{#1}}

\begin{document}

\title{Magnetically tunable Dirac and Weyl fermions in the Zintl materials family}

\author{Anan Bari Sarkar}
\thanks{These authors contributed equally to this work}
\affiliation{Department of Physics, Indian Institute of Technology, Kanpur 208016, India}

\author{Sougata Mardanya}
\thanks{These authors contributed equally to this work}
\affiliation{Department of Physics, Indian Institute of Technology, Kanpur 208016, India}
\affiliation{Department of Physics, National Cheng Kung University, Tainan 701, Taiwan}

\author{Shin-Ming Huang}
\affiliation{Department of Physics, National Sun Yat-sen University, Kaohsiung 80424, Taiwan}

\author{Barun Ghosh}
\affiliation{Department of Physics, Northeastern University, Boston, Massachusetts 02115, USA}

\author{Cheng-Yi Huang}
\affiliation{Department of Physics, Northeastern University, Boston, Massachusetts 02115, USA}
	
\author{Hsin Lin}
\affiliation{Institute of Physics, Academia Sinica, Taipei 11529, Taiwan}

\author{Arun Bansil}
\affiliation{Department of Physics, Northeastern University, Boston, Massachusetts 02115, USA}

\author{Tay-Rong Chang}
\affiliation{Department of Physics, National Cheng Kung University, Tainan 701, Taiwan}
\affiliation{Center for Quantum Frontiers of Research and Technology, Tainan 701, Taiwan}
\affiliation{Physics Division, National Center for Theoretical Sciences, National Taiwan University, Taipei, Taiwan}

\author {Amit Agarwal}
\email{amitag@iitk.ac.in}
\affiliation{Department of Physics, Indian Institute of Technology, Kanpur 208016, India}

\author{Bahadur Singh}
\email {bahadur.singh@tifr.res.in}
\affiliation{Department of Condensed Matter Physics and Materials Science, Tata Institute of Fundamental Research, Mumbai 400005, India}

%\date{\today}
	
\begin{abstract} 

Recent classification efforts encompassing crystalline symmetries have revealed rich possibilities for solid-state systems to support a tapestry of exotic topological states. However, finding materials that realize such states remains a daunting challenge. Here we show how the interplay of topology, symmetry, and magnetism combined with doping and external electric and magnetic field controls can be used to drive the previously unreported SrIn$_2$As$_2$ materials family into a variety of topological phases. Our first-principles calculations and symmetry analysis reveal that SrIn$_2$As$_2$ is a dual topological insulator with $Z_2=(1;000)$ and mirror Chern number $C_M= -1$. Its isostructural and isovalent antiferromagnetic cousin EuIn$_2$As$_2$ is found to be an axion insulator with $Z_4= 2$. The broken time-reversal symmetry via Eu doping in Sr$_{1-x}$Eu$_x$In$_2$As$_2$ results in a higher-order or topological crystalline insulator state depending on the orientation of the magnetic easy axis. We also find that antiferromagnetic EuIn$_2$P$_2$ is a trivial insulator with $Z_4= 0$, and that it undergoes a magnetic field-driven transition to an ideal Weyl fermion or nodal fermion state with $Z_4= 1$ with applied magnetic field. Our study identifies Sr$_{1-x}$Eu$_x$In$_2$(As, P)$_2$ as a new tunable materials platform for investigating the physics and applications of Weyl and nodal fermions in the scaffolding of crystalline and axion insulator states. 

\end{abstract}

\maketitle

\textit{Introduction--} 
Since the discovery of $Z_2$ topological insulators, exploring the role of crystalline symmetries in protecting the non-trivial topology of Bloch Hamiltonians more generally has become a topic of intense research~\cite{RevModPhys.82.3045,RevModPhys.83.1057,bansil2016,Ashvin_RMP}. The interplay of symmetry and topology in the crystalline solids gives birth to many new topological insulators and metals, such as the mirror (glide-mirror)-protected topological crystalline insulators (TCIs)~\cite{Fu_TCI,hsieh2012topological}, rotational-symmetry protected TCIs~\cite{rotation_tci_chen_fang,rotation_tci_TaAs2,Hsu13255}, higher-order topological insulators~\cite{Schindler_2018,HOTI_Bi}, Dirac~\cite{Liu_2014_CdAs,Liu_2014_NaBi,armitage,singh2018}, Weyl~\cite{TaAs_Weyl,TaAs_Ding,Xue1603266}, and unconventional fermion semimetals~\cite{Hourglass_fermion,singh_prl,Bradlynaaf5037}, among other possibilities~\cite{Song2018,TQC_Andrei}. The TCIs are characterized by a unique set of topological invariants which depend only on the dimension and symmetry of the crystalline group~~\cite{rotation_tci_chen_fang,rotation_tci_TaAs2,Hsu13255,Song2018,TQC_Andrei}. The associated topological properties remain robust under symmetry-protecting perturbations and provide materials platforms for developing next-generation low-power-consuming high-speed electronic, optoelectronic, and spintronic devices~\cite{Jungwirth2016,Weyl_opto}. The need for finding new classes of tunable topological materials in order to fill the critical gap in the available materials families is thus clear. 

Manipulation of quantum interactions and topology in the crystalline matrix can provide a promising route for tuning topological properties of materials. The recent discovery of topological magnets in which the nontrivial topology is intertwined with magnetism suggests that certain topological responses could be tuned through external electric and magnetic fields ~\cite{AFM_TI_theory,MnBi2Te4_expt,MBT124_theory,MBT1813,Weyl_magnet,Yin2020}. The topological state in the currently available topological magnets, however, originates from the underlying paramagnetic state and for this reason these materials allow only limited tunability. It is highly desirable, therefore, to find new strategies for identifying tunable topological materials. 

A natural paradigm for discovering new tunable topological materials is to exploit the strong interplay between magnetism and topology in families of isovalent and isostructural materials. Such materials families can offer great flexibility with respect to magnetism, topology, chemical composition, lattice parameters, and transport properties. In particular, spatial and non-spatial symmetries of the parent materials can be relaxed by introducing magnetism with ionic substitutions or crystal structure engineering in a topological non-magnet to drive various different types of topological magnetic orders.~\cite{wang2020topological}. Using this approach as a guiding principle, we identify here the SrIn$_2$As$_2$ class of Zintl materials with the general formula AM$_2$X$_2$ (A = alkali, alkaline earth, or rare-earth, M = post-transition metal; and X = Group IV or V element) in $P6_3/mmc$ space group as a materials family that can support magnetic and nonmagnetic topological phases with substantial flexibility of manipulation through external controls. The Zintl has attracted great interest since its discovery owing to its fascinating electronic, topological, and magnetic properties~\cite{Jiang2006, Mathieu2008, Rauscher_2010, Goforth2008, Guechi_2013, xu2019higher, Zhang2020, Sabin_Eu, riberolles2021, zhang2010eu, shinozaki2021thermoelectric}. 

\begin{figure}[ht!]
\includegraphics[width=0.49\textwidth]{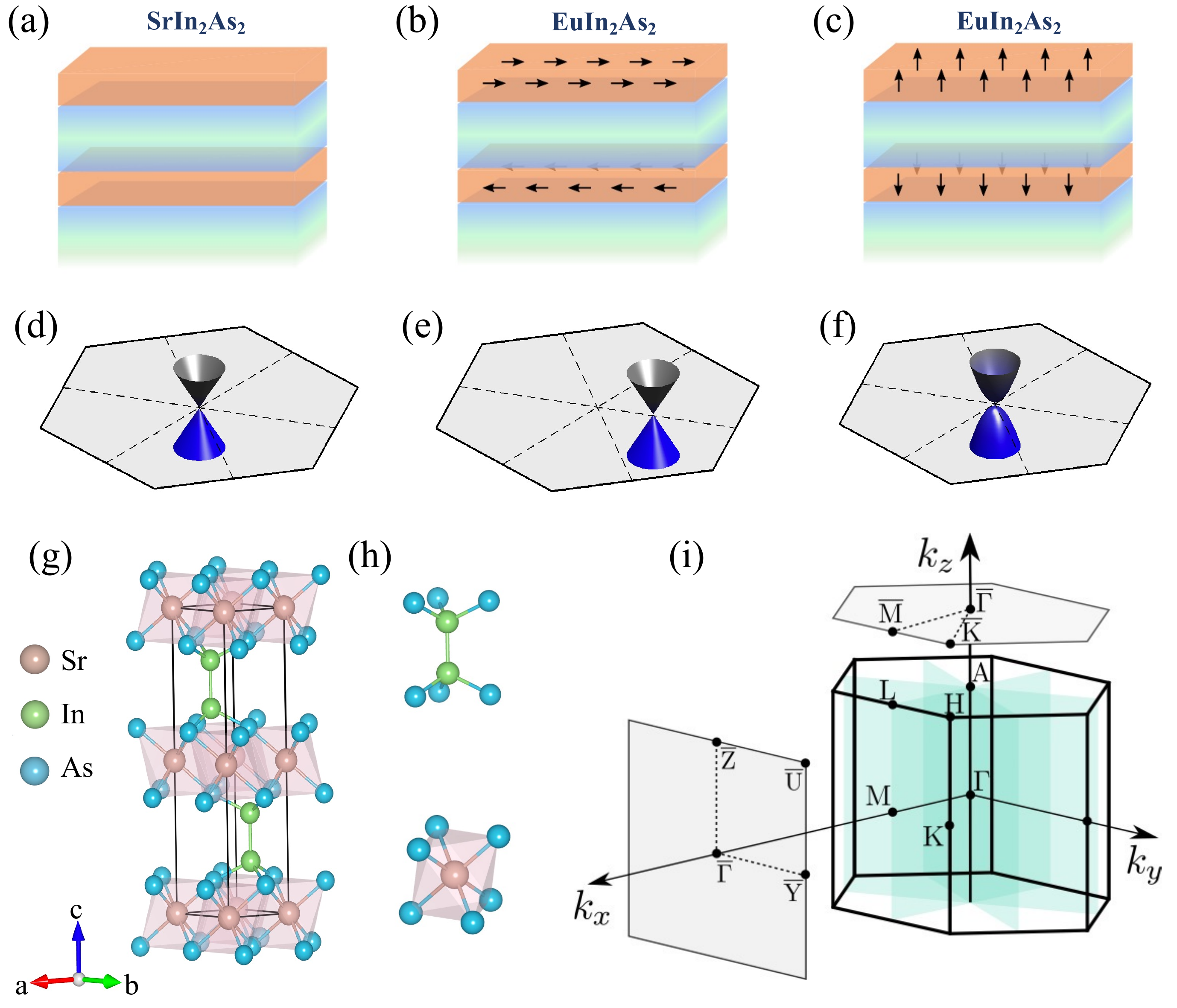} 
\caption{Methodology to design magnetically tunable topological states and crystal structure. (a)-(f) Schematics of the evolution of the topological surface states as we break the mirror and time-reversal symmetries. (a),(d) In the presence of both time-reversal and mirror-symmetry in the nonmagnetic SrIn$_2$As$_2$, the gapless Dirac point appears at the $\overline{\Gamma}$ point. (b),(e) With time-reversal-symmetry broken but mirror-symmetry intact in aAFM EuIn$_2$As$_2$, the gapless Dirac point is pinned to the mirror-symmetric line but shifts away from the $\overline{\Gamma}$ point. (c),(f) On breaking the time-reversal-symmetry and the mirror-symmetry in cAFM EuIn$_2$As$_2$, the surface states become gapped at the $\overline{\Gamma}$ point. (g),(h) The crystal structure of Zintl Eu/SrIn$_2$As$_2$ family showing stacking of the layers of individual atoms along the $c$ axis. (i) Bulk and projected (001) and (100) surface Brillouin zone along with high-symmetry points. The three symmetry equivalent mirror planes $\mathcal{M}_{100}$, $\mathcal{M}_{010}$ and $\mathcal{M}_{110}$ are highlighted with green shading. } 
\label{fig1}
\end{figure}

In Fig.~\ref{fig1}(a)-(f), we demonstrate surface states evolution of three distinct topological phases that could emerge by breaking symmetries with the magnetic orders considering SrIn$_2$As$_2$ and EuIn$_2$As$_2$ as the representative materials. SrIn$_2$As$_2$ is nonmagnetic whereas EuIn$_2$As$_2$ is an A-type antiferromagnetic (AFM) system with Neel temperature of $16K$~\cite{Goforth2008}. SrIn$_2$As$_2$ preserves the time-reversal and spatial symmetries of space group $P6_3/mmc$. It is found to be a dual topological insulator with $Z_2=1$ and mirror Chern number $C_M=-1$ similar to Bi$_2$Te$_3$ class of topological insulators~\cite{Rauch_2014}. The associated Dirac cone states are thus pinned to the $\Gamma$ point (Figs.~\ref{fig1}(a) and (d)). Introducing the AFM order with magnetic moment $m||a$ (aAFM) breaks the time-reversal symmetry but preserves the spatial mirror symmetries (see details below). Thus, aAFM state realizes a topological crystalline insulator in which the Dirac cone states are unpinned from the $\Gamma$ to lie on the mirror invariant line (Figs.~\ref{fig1}(b) and (e)). Considering the out-of-the-plane AFM with magnetic moment $m||c$ (cAFM), the Dirac cone states are fully gapped, realizing an axion or high-order topological insulator state (Figs.~\ref{fig1}(c) and (f)). In this way, a variety of topological states can emanate from the parent materials with relaxing symmetries owing to the magnetic order.

\textit{Methods--}
We performed the first-principle calculations within the framework of the density functional theory (DFT) using the Vienna {\it ab-initio} simulation package (VASP)~\cite{kresse1996efficient, kresse1999ultrasoft}. The generalized gradient approximation (GGA)~\cite{Perdew1996} with Perdew-Burke-Ernzerhof (PBE) parametrization was used for considering the exchange-correlation effects. 
\rpl{We considered  Eu $f$ electrons as valence electrons and adopted the GGA+$U$~\cite{hubbard_U,Anisimov_1997} scheme with $U_{\rm eff}$ = 5.0 eV for Eu $f$ states to include strong electron-correlation effects.}  
A kinetic energy cut-off of 350 eV for the plane-wave basis set and a $\Gamma$-centered $9\times9\times 9$  $k$-mesh~\cite{Monkhorst1976} for bulk Brillouin zone (BZ) integration were used. We consider fully relaxed structural parameters to compute the electronic structure (see SMs for details) ~\cite{Monkhorst1976}. The bulk topological properties and surface spectral functions were calculated using the WannierTools software package~\cite{WU2018405}. 

\textit {Crystal structure and symmetries--} 
The crystal lattice of AM$_2$X$_2$ Zintl compounds is described by the hexagonal space group $P6_3/mmc$ (No. 194) with the atomic arrangement illustrated in Fig.~\ref{fig1}(g)-(h) for SrIn$_2$As$_2$ as an example~\cite{Jiang2006,Rauscher_2010,Guechi_2013}. The Sr, In, and As atoms occupy Wyckoff $2a$, 4$f$, and 4$f$ positions, respectively. This crystal contains covalently bonded two-dimensional (2D) networks of SrIn$_2$As$_2$ along the hexagonal $c-$axis.  Each of SrIn$_2$As$_2$ 2D network contains [In$_2$As$_2$]$^{2-}$ anion layers which are balanced by Sr$^{2+}$ cation layers. The number of valence electrons thus obey the octet rule with filled valence shells ($d^{10}s^2p^6$), leading to a semiconducting or semimetallic ground state. 
\rpl{The Sr atoms can be replaced with other divalent rare-earth magnetic atoms such as Eu to realize magnetic compounds as seen in experiments~\cite{zhang2010eu}.} 
The crystal lattice contains various symmetries which are generated by an inversion $\mathcal{I}$, two-fold rotation axes $\mathcal{C}_{2y}$ and $\mathcal{G}_{2x}=\{\mathcal{C}_{2x}|0 0 \frac{1}{2}\}$, two-fold screw rotation axis $\mathcal{G}_{2z}=\{\mathcal{C}_{2z}| 0 0 \frac{1}{2}\}$, and a three-fold rotation axis $\mathcal{C}_{3z}$ symmetries. These symmetry generators construct twenty-four symmetry operators including a six-fold screw rotation $\{\mathcal{C}_{6z}| 0 0 \frac{1}{2}\}$ ($C_{3z} \otimes \{\mathcal{C}_{2z}| 0 0 \frac{1}{2}\}$), symmetry equivalent vertical mirror $\mathcal{M}_{100}$ and glide-mirror $\{\mathcal{M}_{120}| 0 0 \frac{1}{2}\}$ planes, and a horizontal mirror plane $\{\mathcal{M}_{001}| 0 0 \frac{1}{2}\}$. Figure \ref{fig1}(i) shows the bulk BZ and (001) and (100) surface projected BZs where various high-symmetry points marked.\\

%% Crystal structure

\begin{figure}[ht!]
\includegraphics[width=0.49\textwidth]{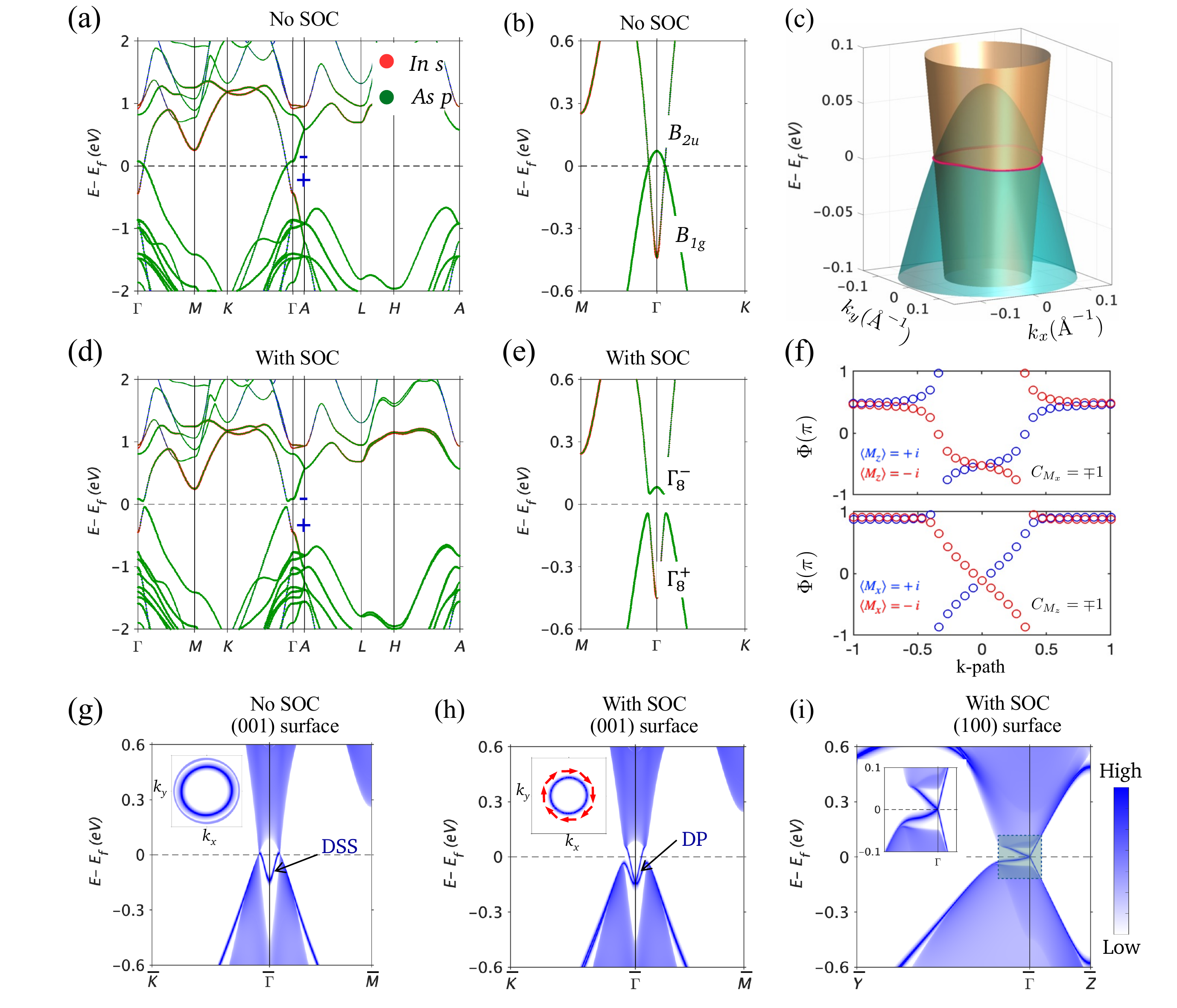} 
\caption{Topological band structure of nonmagnetic SrIn$_2$As$_2$. (a) Calculated bulk band structure without spin-orbit coupling (SOC). The parity eigenvalues ($\pm$) and the In $s$ (red) and As $p$ (green) orbitals are marked. (b) Closeup of the bands along the $\Gamma-K$ and $\Gamma-M$ directions in the $k_z=0$ plane. Band crossings between $B_{1g}$ and $B_{2u}$ states with opposite $M_z=0$ mirror-eigenvalue are evident. (c) $E-k_x-k_y$ rendition of the valence and conduction bands on the $k_z=0$ plane. The nodal line is shown with the red line. (d),(e) Same as (a),(b) but with the inclusion of SOC. A hybridization gap between the valence and conduction bands is seen. (f) Evolution of the Wannier charge center on the $k_z=0$ and $k_x=0$ planes. The non-zero mirror Chern number confirms the mirror TCI phase in SrIn$_2$As$_2$. (g)-(h) Band structure of (001) surface without and with SOC. The inset shows the fermi band contours associated with the calculated surface states. (i) Surface band structure of (100) surface with SOC. Symmetry-enforced anisotropic Dirac cone is highlighted in the inset.}
\label{fig2}
\end{figure}

\textit {Nonmagnetic topological states--}
Topological states can be identified by an inverted band structure owing to the high spin-orbit coupling (SOC) or crystal field effects. The bulk band structure and band symmetries of SrIn$_2$As$_2$ in the absence of SOC is shown in Fig.~\ref{fig2}(a). It constitutes a band inversion between $B_{1g}$ and $B_{2u}$ states at the $\Gamma$ point (Fig.~\ref{fig2}(b)). The orbital analysis illustrates that the $B_{1g}$ states are derived from the In $s$ orbitals and lie below the As $p$ orbitals constituted $B_{2u}$ states. This band order remains normal at other high-symmetry points in BZ. Away from the $\Gamma$ point, $B_{1g}$ and $B_{2u}$ states are seen to cross at the $\Gamma-M$ and $\Gamma-K$ lines at $k= 0.0035$ $\mathrm{\AA}^{-1}$ and $0.0049$ $\mathrm{\AA}^{-1}$, respectively. A full BZ exploration shows that these band crossings trace a topological Dirac nodal line on the $k_z=0$ plane as shown in Fig.~\ref{fig2}(c). SrIn$_2$As$_2$ thus realizes a band inversion topological nodal line semimetal without SOC. The robustness of this topological state is further verified via parallel computations using the hybrid density functional given the possible underestimation of the bandgap in GGA functional (see supplementary materials (SMs) for details). 

%% Nonmagnetic state

The band structure in the presence of SOC (Fig.~\ref{fig2}(d)) shows a continuous gap between the $\Gamma_8^+{}$ and $\Gamma_8^-{}$ states at the Fermi level. Figure \ref{fig2}(e) shows the closeup of the bands along the $\Gamma-M$ and $\Gamma-K$ directions and highlights band anti-crossings, showing that the SOC drives the system into an inverted insulator state. Since the system preserves inversion symmetry $\mathcal{I}$, the parity analysis at the eight time-reversal invariant momentum (TRIM) points gives $Z_2=(1;000)$. On the other hand, the $\mathcal{M}_{001}$ and $\mathcal{M}_{110}$ mirror Chern numbers on $k_z=0$ and $k_{y} = 0$ planes are calculated as -1 and -1, respectively, from the Wannier charge centers evolution as shown in Fig.~\ref{fig2}(f). SrIn$_2$As$_2$ is thus a dual topological insulator.

The nontrivial topological properties of bulk manifest on the surface. From the (001) surface energy dispersion in Fig.~\ref{fig2}(h), a topological Dirac cone state with the Dirac point at the $\overline{\Gamma}$ point is resolved. The associated constant energy contours are circular with isotropic energy dispersion near the Dirac node and develop a hexagonal shape as one moves away from the Dirac point. The spin-texture is helical with a left-handed chirality for the upper Dirac cone as shown in the inset of Fig.~\ref{fig2}(h). This surface state is protected by both the time-reversal symmetry and $\mathcal{M}_{001}$ mirror symmetry. We also get a similar protected topological state on the low-symmetry (100) side surface. However, the Dirac cone on the (100) surface is located well within the bulk energy gap [see Fig.~\ref{fig2}(i)]. It has an anisotropic energy dispersion, indicating that the surface Dirac fermions have different Fermi velocities along different directions.

\begin{figure}[ht!]
\includegraphics[width=0.49\textwidth]{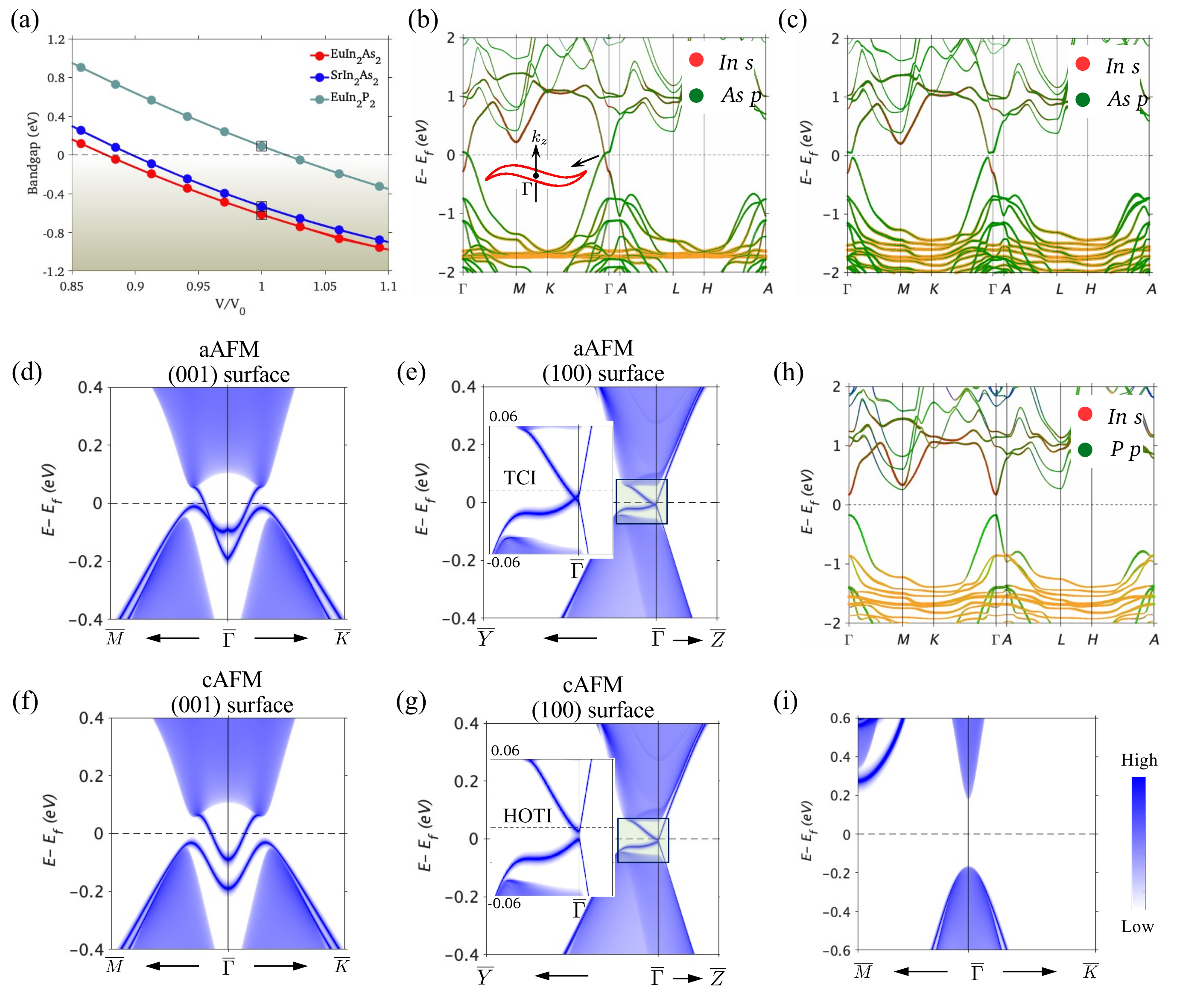} 
 \caption{Topological states and magnetic tunability of antiferromagnetic EuIn$_2$(As, P)$_2$. (a) Variation of the bandgap as a function of relative volume at the $\Gamma$ point for AFM EuIn$_2$(As, P)$_2$ and non-magnetic SrIn$_2$As$_2$. Open markers at $V/V_0=1$ show the bandgap of their pristine states. A topological phase transition from the nontrivial to trivial phase can be achieved by decreasing the volume. Calculated band structure of AFM EuIn$_2$As$_2$ (b) without and (c) with SOC. The crossings in (b) form a nodal line unhooked from $k_z=0$ plane due to breaking of $\mathcal{M}_{001}$ mirror plane symmetry. A hybridization gap opens up at the crossing points and an Axion insulator state with $Z_4=2$ develops with SOC. The calculated surface band structure of aAFM EuIn$_2$As$_2$ for (d) (001) and (e) (100) surfaces. The mirror symmetry-protected Dirac cones states are seen inside the bulk gap. (f)-(g) same as (d)-(e) but for cAFM order. The topological surface states are now gapped with a magnetic gap at the $\overline{\Gamma}$ point. (h)-(i) Calculated (h) bulk and (i) (001) surface band structure of aAFM EuIn$_2$P$_2$. A trivial insulating state with $Z_4=0$ is resolved.}
\label{fig3}
\end{figure}

\textit {Magnetic topological states--}
We now consider the topological state of EuIn$_2$As$_2$ which is a magnetic counterpart of SrIn$_2$As$_2$. Notably, EuIn$_2$(As, P)$_2$ is isostructural and isovalent to SrIn$_2$As$_2$ but realizes an AFM state below $~16$K~\cite{Rauscher_2010}. EuIn$_2$As$_2$ shows immense potential for realizing Axion insulator state and for this reason it has attracted attention recently ~\cite{xu2019higher, Zhang2020,Sabin_Eu,riberolles2021}. 
\rpl{The neutron diffraction experiments unfold a low-symmetry helical antiferromagnetic order for EuIn$_2$As$_2$ which is further amenable for realizing various topological states under external magnetic field~\cite{riberolles2021}. The experimentally saturated magnetic moment for Eu ions is $\sim 7.0~\mu_{\text{B}}$ that is close to the magnetic moment of $f^7$ state and our calculated value of $6.90~\mu_{\text{B}}$.} 
This indicates that Eu oxidation state is +2 similar to the oxidation state of Sr. A tunable Sr$_{1-x}$Eu$_x$In$_2$(As, P)$_2$ system thus can be achievable for realizing distinct topological magnetic states ~\cite{zhang2010eu,shinozaki2021thermoelectric}.
The dependence of the bandgap on the lattice constant (Fig.~\ref{fig3}(a)) shows that EuIn$_2$As$_2$ has an intrinsically inverted band structure whereas EuIn$_2$P$_2$ has a trivial band structure. 

Figure~\ref{fig3} shows the topological states of antiferromagnetic EuIn$_2$(As, P)$_2$. We consider topological states originating from the formation of commensurate AFM order in EuIn$_2$As$_2$ with the intra-layer ferromagnetic coupling and inter-layer anti-ferromagnetic coupling between the Eu spins. The calculated energy for the aAFM is nearly the same as cAFM, making both these states energetically favorable. This nearly degenerate energy between different magnetic orders suggests that external magnetic tuning of the topological state in EuIn$_2$As$_2$ would be possible~\cite{Zhang2020,riberolles2021}.

%% Magnetic state

\rpl{We emphasize that considering the AFM order reduces the nonmagnetic $D_{6h}$ point-group symmetry to either $D_{2h}$ or $D_{3d}$ depending on the magnetic configuration of the AFM state.} Figure \ref{fig3}(b) shows the band structure of AFM EuIn$_2$As$_2$ without SOC. It retains the characteristic inverted band features seen in SrIn$_2$As$_2$ along with the Dirac nodal line in the BZ. 
\rpl{Specifically, the states close to the Fermi level are comprised of In $s$ and As $p$ orbitals, whereas Eu $f$ states provide the necessary exchange coupling for inducing the magnetism in the system. Often strong electron correlations in Eu $f$ states affect band ordering and thus the topological state of materials ~\cite{bendavid2014first, prokofiev1996periodicity, yu2020machine}. Our analysis by varying Hubbard $U_{\text{eff}}$ in GGA+U in parallel with hybrid functional calculations reveals that states close to Fermi level and the topological phase of EuIn$_2$As$_2$ do not depend on the choice of U parameter (see SMs). }
However, magnetism breaks the horizontal $\mathcal{M}_{001}$ mirror plane, and thus unlike non-magnetic SrIn$_2$As$_2$, the nodal-line in EuIn$_2$As$_2$ for a particular spin is unhooked from the $k_z=0$ plane and wriggles in the momentum space around the $\Gamma$ point with small $k_z$ dispersion (Fig.~\ref{fig3}(b)).  Inclusion of SOC in computations develops a gap at the band crossing points, separating valence and conductions bands in both the aAFM and cAFM states (Fig.~\ref{fig3}(c)). Since both the AFM states preserve inversion $\mathcal{I}$ symmetry, we access their topological character by calculating the parity-based $Z_4$ topological invariant. The calculated $Z_4$ is 2, indicating that EuIn$_2$As$_2$ realizes an Axion insulator state. Notably, despite a nonzero $Z_4$ invariant, the topological state of the system can further be classified based on the magnetic group symmetries as discussed below.

The aAFM state of EuIn$_2$As$_2$ preserves $\mathcal{M}_{001}$ and $\mathcal{M}_{100}$ mirror planes. The calculated mirror Chern number on the $k_z=0$ and $k_{y} = 0$ planes is -1, similar to SrIn$_2$As$_2$. EuIn$_2$As$_2$ is thus a mirror topological crystalline insulator with aAFM order. This is further manifested in the surface spectrum on the (001) and (100) planes where Dirac cones are seen along the mirror invariant lines. Specifically, the mirror-protected Dirac cone is located on the $\overline{\Gamma}-\overline{M}$ line of the (001) plane and $\overline{\Gamma}-\overline{Y}$ line of the (100) plane as depicted in Figs. $\ref{fig3}$(d) and $\ref{fig3}$(e). On considering cAFM state, both the $\mathcal{M}_{001}$ and $\mathcal{M}_{100}$ mirror planes are broken. There is no symmetry-protected Dirac cone state found on either surface. Instead, a magnetic gap develops at the surface Dirac node crossings (see Fig.~\ref{fig3}(f) and (g)). The system thus realizes a higher-order topological insulator with conducting hinge states~\cite{Ezawa}. Figures~\ref{fig3}(h) and \ref{fig3}(i) show the bulk and surface band structure of AFM EuIn$_2$P$_2$. In contrast to EuIn$_2$As$_2$, EuIn$_2$P$_2$ exhibits a trivial state with $Z_4=0$. 
\rpl{This trivial state may arise due to strong carrier confinement. EuIn$_2$P$_2$ has a relatively small unit cell volume (267.03 \AA$^3$) than EuIn$_2$As$_2$ (290.77 \AA$^3$) which could induce strong carrier confinement effect. Such effects reduce the bandwidth for both the valence and conduction bands, thereby avoiding the band inversion in the vicinity of the Fermi level. }

To showcase the novel physics that emerges in the presence of an external magnetic field, we now discuss the topological states in ferromagnetic (FM) EuIn$_2$(As, P)$_2$. Our total energy calculations show that the FM state is $\sim$2 (3)  meV/unit cell higher than the AFM state of EuIn$_2$As$_2$ (EuIn$_2$P$_2$). Notably, we have considered the commensurate FM orders with the magnetic moment $m||a$ (aFM) and magnetic moment $m||c$ (cFM). Both the aFM and cFM possess nearly degenerate energy similar to the AFM states. It is thus possible that FM order in these materials becomes energetically favorable under a small external magnetic field. For FM EuIn$_2$As$_2$, we find that the bulk band inversion stays robust for both the FM orders and the system realizes an Axion insulator state with $Z_4=2$ (see SMs).  

%% Weyl semimetal
\begin{figure}[ht!]
 \includegraphics[width=0.49\textwidth]{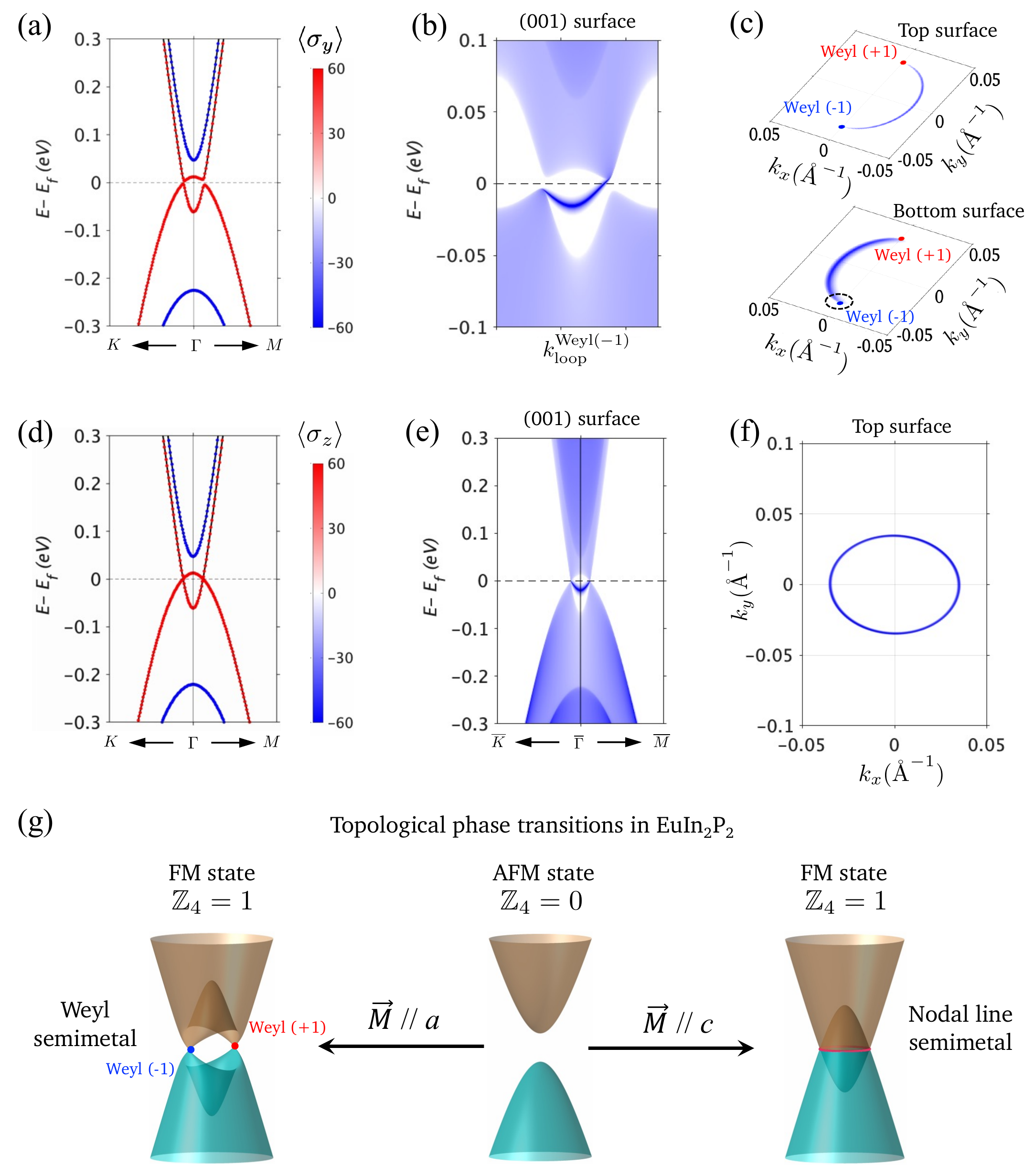} 
 \caption{Ferromagnetic tunability of topological states of EuIn$_2$P$_2$. (a) Calculated spin-resolved bulk band structure of A-type ferromagnetic (aFM) configuration. A single spin-band-inversion drives aFM phase into an ideal ferromagnetic Weyl semimetal with a pair of Weyl nodes on the $k_x-k_y$ plane. (b) The surface band structure calculated along the circular path shown in (c) on the (001) surface enclosing a Weyl point. The chiral surface states are seen connecting the projected bulk valance and conduction bands. (c) The Fermi-arc surface states, connecting projected Weyl nodes on the top and bottom surfaces. (d) Spin-resolved bulk (f) (001) surface band structure of C-type ferromagnetic (cFM) configuration. The presence of $\mathcal{M}_{001}$ mirror plane in cFM configuration protects the crossing bands to hybridize, realizing a nodal line with drumhead surface state. (g) The associated Fermi band contours showing the drumhead surface states. (h) Schematic representation of the trivial insulator ($Z_4=0$) to a Weyl semimetal or nodal line semimetal ($Z_4=1$) transition in EuIn$_2$P$_2$ with applied external magnetic field. } 
\label{fig4}
\end{figure}

Figure~\ref{fig4} illustrates the bulk and surface band structure of FM EuIn$_2$P$_2$. Due to the large exchange coupling of Eu $f$ states, it shows a single spin-band inversion in both aFM and cFM orders. Our parity-based topological indicator is obtained to be $Z_4 = 1$, indicating that the system is a nontrivial half-semimetal. Since this type of topological semimetal state of a magnetic system depends on crystalline symmetries, one can realize both the time-reversal broken Weyl semimetal or nodal line semimetal state. On considering aFM state, we find that aFM state breaks $\mathcal{M}_{100}$ mirror plane symmetry, and the system realizes an ideal Weyl semimetal with two Weyl nodes in the bulk BZ, as shown in Fig.~\ref{fig4}(a). The Weyl nodes are located on the $k_z= 0$ plane at (0.000, 0.0355, 0.0) $\mathrm{\AA}^{-1}$ with chirality $+1$ and at (0.000, -0.0355, 0.0) $\mathrm{\AA}^{-1}$ with chirality $-1$ (see SMs). The surface band structure and the surface Fermi arcs connecting the projected Weyl nodes on the (001) surface are shown in Figs. ~\ref{fig4}(b)-(c). Figures~\ref{fig4}(d)-(f) show the bulk and surface electronic spectrum of the cFM state. Since the cFM state preserves the $\mathcal{M}_{001}$ mirror symmetry, the two crossing bands in Fig.~\ref{fig4}(d) stay robust against band hybridization, thereby realizing a single Weyl nodal line on the $k_z= 0$ plane. The associated nontrivial drumhead surface states and Fermi band contours are shown in Figs.~\ref{fig4}(e)-(f). We emphasize an FM state can be considered as a fully polarized final state of an initial AFM state. It is thus possible to realize the aforementioned ideal topological half-semimetal states in EuIn$_2$P$_2$. This is shown schematically in Fig.~\ref{fig4}(g). A magnetic field applied parallel to the $y$ axis will drive it into a nearly ideal Weyl semimetal whereas a magnetic field applied parallel to the $z$ axis will drive it into a Weyl nodal-line semimetal state.    

\rpl{
\textit{$k\cdot p$ model analysis--} The preceding analysis reveals that the band inversion happens between In $s$ and As $p$ states at $\Gamma$ in both the non-magnetic and magnetic compounds. The role of magnetism is to reduce the symmetries and drive the system into various TCI states. We can thus consider the Hamiltonian of the nonmagnetic state and regard magnetism as perturbation allowed by the magnetic symmetries so that $H=H_{0}+\Delta H$, where $H_{0}$ is non-magnetic state Hamiltonian whereas $\Delta H$ originates from the magnetic order. The irreducible representations of two bands at $\Gamma$ close to the Fermi level are $B_{1g}$ and $B_{2u}$ of symmetry group $D_{6h}$ as shown in Fig.~\ref{fig1}(b). Based on the theory of invariants, the low-energy effective Hamiltonian can be described as  
\begin{equation}
    H_0(\vec{k})=\epsilon(\vec{k}) \tau_0 \sigma_{0} +  A_{1} \tau_x\left( k_x \sigma_y - k_y \sigma_x\right)
    +A_{2} k_z \tau_{y} \sigma_{0} + M(\vec{k}) \tau_z \sigma_{0},
\end{equation}
where $\epsilon(\vec{k}) = \epsilon_0+\epsilon_1(k_x^2+k_y^2)+\epsilon_2 k_z^2$ and $M(\vec{k})=M_0-\beta_1(k_x^2+k_y^2)-\beta_2 k_z^2 $. $\tau$ and $\sigma$ are the Pauli spin matrices in the orbital and spin space, respectively, and $M_0$, $\beta_1$ and, $\beta_2>0$ (see SM for details). When SOC is off, $A_{1}=0$ which leads to a nodal ring in the $k_z=0$ plane described with $k_x^2+k_y^2 = M_0/\beta_1$.  
A cAFM order that breaks $\mathcal{T}$, $\mathcal{C}_{2y}$, and $\mathcal{G}_{2z}$ while preserving $\mathcal{TC}_{2y}=\tau_z \sigma_{0} \mathcal{K}$ and $\mathcal{C}_{2y}\mathcal{G}_{2z}=\mathcal{G}_{2x}$ symmetry operations gives  
\begin{equation}
    \Delta H_{\text{cAFM}}(\vec{k})=
     \left( B_1\tau_0 +B_2 \tau_z\right) \left[ \left( k_x^2-k_y^2 \right) \sigma_x - 2k_x k_y \sigma_y\right].
\end{equation} 
A aAFM order that breaks $\mathcal{T}$ and $\mathcal{C}_{3z}$ gives  
\begin{equation}
   \Delta H_{\text{aAFM}}(\vec{k})= 
    D_0 \tau_y \left( k_x \sigma_y - \delta_{\varepsilon} k_y \sigma_x \right).  
\end{equation}
The ferromagnetic cFM and aFM phases can come with $\Delta H_{\text{cFM}}=\lambda_{c} \tau_z \sigma_z$ and $\Delta H_{\text{aFM}}=\lambda_{a} \tau_z \sigma_y$, respectively.}

\textit {Summary--}
\begin{table*}[t!]
\caption{Summary of topological states in SrIn$_2$As$_2$, EuIn$_2$As$_2$, and EuIn$_2$P$_2$ Zintl materials for different commensurate magnetic configurations. NM stands for nonmagnetic, aAFM (cAFM) denotes antiferromagnetic state with magnetic moment $m||a$ ($m||c$), and aFM (cFM) represents ferromagnetic state with magnetic moment $m||a$ ($m||c$). TI, TCI, AI, and NI stand for topological insulator, topological crystalline insulator, Axion insulator, and normal insulator, respectively. WSM and NLSM denote Weyl semimetal and nodal line semimetal, respectively. See text for more details. }
\centering
\begin{tabular}{c c c c c c}
\hline \hline 
Compound           				& Magnetic state  &	 Space group       				& Topological invariant & Topological state	\\
\hline \\
\multirow{1}{*} {SrIn$_2$As$_2$ }   &  	NM             & 	$P6_3/mmc$  				  	&      $Z_2 = 1$         		&   TI $\&$ TCI     	\\ 
\multirow{4}{*} {EuIn$_2$As$_2$ }  &    aAFM           &	$Cmcm$    			& 	$Z_4 = 2$        	 	& TCI    	  		\\  
  							 &    cAFM           &	$P6_{3^{\prime}}/m^{\prime}m^{\prime}c$       						& 	$Z_4 = 2$        		& AI        			\\
                                				&    aFM             &	$Cmc^{\prime}m^{\prime}$  			& 	$Z_4 = 2$        		& TCI 			\\
                                    				&    cFM             & 	$P6_3/mm^{\prime}c^{\prime}$            						& 	$Z_4 = 2$         		& AI				\\  
\multirow{4}{*} {EuIn$_2$P$_2$ }    &   aAFM          	& 	$Cmcm$         						& 	$Z_4 = 0$         		& NI				\\
                                   				&   cAFM           & 	$P6_{3^{\prime}}/m^{\prime}m^{\prime}c$         						& 	$Z_4 = 0$         		& NI 			\\ 
                                 			  	&   aFM              & 	$Cmc^{\prime}m^{\prime}$         						& 	$Z_4 = 1$    		& WSM			 \\
                                   				&  cFM               & $P6_3/mm^{\prime}c^{\prime}$   				& 	$Z_4 = 1$         		& NLSM			\\ 
 \hline  \hline
\end{tabular} \label{T1:bulk}
\end{table*} 
We have shown that the interplay of topology, symmetry, and magnetism in the experimentally realized SrIn$_2$As$_2$ materials class can enable access to a multitude of topological states through the lowering of symmetry via various magnetic configurations. SrIn$_2$As$_2$, a hexagonal Zintl compound, is found to be a dual topological insulator with nontrivial $Z_2 =1$ and mirror Chern number $C_M=-1$. Its isostructural and isovalent magnetic cousin EuIn$_2$ (As, P)$_2$ supports a variety of topological states. Remarkably, we find that EuIn$_2$As$_2$ is an axion insulator with $Z_4=2$, which can also realize a mirror topological magnetic insulator for aAFM (magnetic moment $m||a$) and a higher-order topological insulator for cAFM ($m||c$). In contrast, EuIn$_2$P$_2$ is a robust AFM trivial insulator in its pristine form and undergoes a magnetic field-driven transition to an ideal ferromagnetic Weyl point ($m||a$) and Weyl nodal line ($m||c$) half-semimetal with $Z_4=1$.  In conjunction with the bulk topological states, we also obtain a the variety of surface states viz. gapless or gapful, pinned or unpinned at the $\Gamma$ point, see Table~\ref{T1:bulk} for a summary of topological states in (Sr,~Eu)In$_2$(As, P)$_2$. Since Sr and Eu are isovalent, Sr$_{1-x}$Eu$_x$In$_2$(As, P)$_2$ provides a material that can be tuned systematically between various topological orders through its inherent magnetism and external stimuli. 
\rpl{A topological phase transition from an Axion insulator ($Z_4=2$) to a trivial insulator ($Z_4=0$) can be realized in EuIn$_2$As$_{2-x}$P$_{x}$ by varying the doping concentration $x$ as discussed in SMs.} 
Our results thus highlight that the SrIn$_2$As$_2$ class of topological insulators in the Zintl family are a promising materials platform for investigating the interplay of topology, symmetry, and magnetism and for device applications. 

\textit {Acknowledgements--} A. B. S. acknowledges IIT Kanpur for providing a Senior Research Fellowship. A.B.S. and A.A. thank the CC-IITK for providing the HPC facility. A.A. acknowledge funding from Science Education and Research Board (SERB) and Department of Science and Technology (DST), Government of India. T.-R.C. was supported by the Young Scholar Fellowship Program from the Ministry of Science and Technology (MOST) in Taiwan, under a MOST grant for the Columbus Program MOST110-2636-M-006-016, NCKU, Taiwan, and National Center for Theoretical Sciences, Taiwan. Work at NCKU was supported by the MOST, Taiwan, under grant MOST107-2627-E-006-001 and Higher Education Sprout Project, Ministry of Education to the Headquarters of University Advancement at NCKU. S.M.H. is supported by the MOST-AFOSR program on Topological and Nanostructured Materials, Grant No. 110-2124-M-110-002-MY3. H.L. acknowledges the support by the Ministry of Science and Technology (MOST) in Taiwan under grant number MOST 109-2112-M-001-014-MY3. The work at Northeastern University was supported by the Air Force Office of Scientific Research under award number FA955-20-1-0322 and benefited from the computational resources of Northeastern University’s Advanced Scientific Computation Center (ASCC) and the Discovery Cluster. The work at TIFR Mumbai was supported by the Department of Atomic Energy of the Government of India under Project No. 12-R$\&$D-TFR- 5.10-0100.

\bibliography{Ref_Eu}

\end{document}